\documentstyle[aps,12pt,amssymb,epsfig]{revtex}

\begin{document}

\baselineskip=0.75cm
\newcommand{\ini}{\begin{equation}}
\newcommand{\fin}{\end{equation}}
\newcommand{\inir}{\begin{eqnarray}}
\newcommand{\finr}{\end{eqnarray}}
\newcommand{\inif}{\begin{figure}}
\newcommand{\finf}{\end{figure}}
\newcommand{\bc}{\begin{center}}
\newcommand{\ec}{\end{center}}

\def\ol{\overline}
\def\pa{\partial}
\def\ra{\rightarrow}
\def\ts{\times}
\def\df{\dotfill}
\def\bs{\backslash}
\def\dg{\dagger}

$~$

\hfill DSF-35/2001

\vspace{1 cm}

\centerline{{\bf BARYON ASYMMETRY AND MASS MATRICES}}

\vspace{1 cm}

\centerline{\large{D. Falcone}}

\vspace{1 cm}

\centerline{Dipartimento di Scienze Fisiche, Universit\`a di Napoli,}
\centerline{Complesso di Monte S. Angelo, Via Cintia, Napoli, Italy}

\vspace{1 cm}

\begin{abstract}

\noindent
In the framework of the baryogenesis via leptogenesis mechanism we study
the link between the amount of baryon asymmetry and neutrino mass
matrices. In particular, neglecting phases,
we find that if the Dirac neutrino mass matrix is related to the up quark
mass matrix, the baryon asymmetry is about three orders smaller than the
required value.
If the Dirac neutrino mass matrix is related to the down quark or the charged
lepton mass matrix, the baryon asymmetry is about two orders smaller than
the required value. In order to get a sufficient amount of baryon asymmetry
we need a more moderate hierarchy in the Dirac neutrino mass matrix.

\end{abstract}

\newpage

\section{Introduction}

The origin and amount of baryon asymmetry in our universe depends on both
cosmological features and particle-physics properties.
Sakharov \cite{sakh} discovered
that three conditions must be realized in order to obtain a baryon asymmetry:
baryon number ($B$) violation, charge conjugation ($C$) and combined charge
conjugation and parity ($CP$) violations, and finally out-of-equilibrium
dynamics. The first two conditions come out from particle physics, while the
third one is usually provided by the expansion of the universe.

The typical particle-physics theory in which these conditions can be realized
is the grand unified theory \cite{yoshi},
where $B$ violation is a key signature. However,
this approach has some shortcomings.
In fact,  electroweak sphaleron processes \cite{krs} could wash out a previously
created baryon asymmetry \cite{rt}. Moreover, unified theories are not yet
confirmed by proton decay.

Another approach is based on the production of the baryon asymmetry at the
electroweak scale by sphaleron processes \cite{krs}. However, the standard model
realization fails, due to the lower mass limit of the Higgs boson, and the
supersymmetric version is in the corner of the parameter space \cite{rt}.

Then, a good alternative is the baryogenesis via leptogenesis mechanism
\cite{fy,luty}, based on the out-of-equilibrium decay of heavy right-handed Majorana
neutrinos, which generates a lepton asymmetry to be partially converted into a
baryon asymmetry by the sphaleron processes. Indeed, the existence of very heavy
neutrinos can account for the smallness of effective neutrino mass by means
of the seesaw mechanism \cite{ss}.

In this paper we explore the link between the amount of baryon asymmetry
generated in the baryogenesis via leptogenesis mechanism and quark-lepton
mass matrices. We assume large mixing of solar neutrinos and relate the Dirac
neutrino mass matrix to the up quark or down quark mass matrix.

The present article is similar to an update of Ref.\cite{bb}. In fact, we use
double (atmospheric and solar) large lepton mixing, which is favoured by
recent data, instead of single (atmospheric) large lepton mixing, and quark
mass matrices from the recent Ref.\cite{dv}, instead of the mass matrices
from Ref.\cite{rrr}.

In section II the baryogenesis via leptogenesis is introduced. In section III
the link between baryon asymmetry and mass matrices is studied,
and finally, in section IV, we give our conclusion.

\section{Baryogenesis via leptogenesis}

A baryon asymmetry can be generated from a lepton asymmetry, due to electroweak
sphaleron processes \cite{fy}. The lepton asymmetry is produced by the
decay of heavy right-handed neutrinos, which are Majorana particles and
therefore their mass terms violate lepton number ($L$).
The sphalerons, which violate $B+L$ but conserve $B-L$, convert part of this
lepton asymmetry into a baryon asymmetry.
The $CP$ violation is present because of complex Yukawa
couplings of right-handed (singlet) neutrinos with the left-handed lepton
doublet and the Higgs doublet. These interactions generate the decay of heavy
neutrinos and also the Dirac
neutrino mass matrix $M_{\nu}$ through the vacuum expectation value (VEV) of the
Higgs doublet.

The baryon asymmetry can be defined as \cite{kt}
\ini
Y_B=\frac{n_B-n_{\ol{B}}}{{s}}=\frac{n_B-n_{\ol{B}}}{7 n_{\gamma}}=
\frac{\eta}{7},
\fin
where $n_B$, $n_{\ol{B}}$, $n_{\gamma}$ are number densities,
$s$ is the entropy density, and $\eta$ is the baryon-to-photon ratio.
The master formula for the calculation of the baryon asymmetry
in the baryogenesis via leptogenesis mechanism is given by
\ini
Y_B \simeq \frac{1}{2} \frac{1}{g^*} ~d~\epsilon_1,
\fin
where
\ini
\epsilon_1 \simeq \frac{3}{16 \pi}
\left[\frac{\text{Im}[(y_D^{\dg} y_D)_{12}]^2}{(y_D^{\dg} y_D)_{11}}
\frac{M_1}{M_2}+\frac{\text{Im}[(y_D^{\dg} y_D)_{13}]^2}{(y_D^{\dg} y_D)_{11}}
\frac{M_1}{M_3} \right]
\fin
is a $CP$ violating asymmetry, $d$ is a dilution factor to be discussed
in the following, and $g^* \simeq 100$ counts the light degrees of
freedom in the theory (see for example Ref.\cite{ft} and references therein).
The $CP$ violating asymmetry comes out from the interference between the
tree level and one loop graphs in the out-of-equilibrium decay of the
lightest heavy neutrino. This lightest neutrino is in equilibrium during the
decays of the two heavier ones, washing out the lepton asymmetry generated
by them. The Yukawa matrices $y_{\nu}$ are given by
$M_{\nu}/v$, where $v \simeq m_t$ is the VEV of the Higgs doublet. Matrices $y_D$ are
obtained by $y_D =y_{\nu} U_R$, where the unitary matrix $U_R$ diagonalizes
the mass matrix of heavy neutrinos,
$M_R$, with three eigenvalues $M_1 < M_2 < M_3$. The factor $1/2$ in Eqn.(2)
indicates the part of the lepton asymmetry which is converted into a baryon
asymmetry \cite{ht}.

The dilution factor should be calculated by solving the Boltzmann equations
of the system. It includes the effect of the decay width of the lightest
heavy neutrino, as well as the wash out effect of lepton number violating
scatterings.
In Ref.\cite{bp} it is shown that the dilution depends mostly on the
mass parameter
\ini
\tilde{m}_1 = \frac{(M_D^{\dg} M_D)_{11}}{M_1},
\fin
with $M_D=y_D v$,
and for high $\tilde{m}_1$ some dependence on $M_1$ also appears.
Minor dilution, $d$ of order $10^{-1}$, is obtained for
$10^{-5} < \tilde{m}_1 < 10^{-2}$ eV. If $\tilde{m}_1$ is too low, the Yukawa
couplings are too small to produce a sufficient number of heavy neutrinos at
high temperature, while if $\tilde{m}_1$ is too large, the wash out effect is
too strong and destroys the generated asymmetry.
See also the discussion in Ref.\cite{hk}.
In the present paper, emphasis is given to the possible dependence of the
baryon asymmetry on neutrino mass matrices, without a detailed study of
the dilution.

\section{Leptogenesis and mass matrices}

In this section we explore the link between the baryon asymmetry and lepton
mass matrices. We use symmetric phenomenologically allowed forms
for the quark mass matrices with five and four texture zeros
\cite{dv,cf}, and get neutrino mass matrices by
relating them to the up quark mass matrix $M_u$ or the down quark mass matrix
$M_d$.
The charged lepton mass matrix $M_e$ is always related to the down quark
mass matrix. This is suggested by unified and left-right models \cite{gn,fal1}. 
As a matter of fact, the baryogenesis via leptogenesis mechanism is active
also within such theories (see the review \cite{pil} and the two papers in
Ref.\cite{cfl}). However, we may also assume the quark-lepton mass relations
as phenomenological inputs within the standard model with heavy right-handed
neutrinos, in such a way to avoid possible problems with proton decay
\cite{pro}.

We consider only the large mixing solution to the solar
neutrino problem, which is favoured by recent analyses. In Ref.\cite{fal2}
it is shown that the double large lepton mixing
with the hierarchy $m_1 \ll m_2 \ll m_3$ for the light (effective) Majorana
neutrinos leads to the approximate democratic form
\ini
M_L^{-1} \sim \left( \begin{array}{ccc}
1 & 1 & 1 \\ 1 & 1 & 1 \\ 1 & 1 & 1
\end{array} \right) \frac{1}{m_1},
\fin
where $M_L$ is the effective neutrino mass matrix. We obtain the mass matrix
of heavy neutrinos by means of the inverse seesaw formula
\ini
M_R \simeq M_{\nu} M_L^{-1} M_{\nu}.
\fin
In this way, $M_R$ is obtained by following a kind of bottom-up approach.
The Dirac neutrino mass matrix comes from a theoretical or phenomenological
hint, and then the heavy neutrino mass matrix is inferred, through the inverse
seesaw formula, from the effective neutrino mass matrix.

Charged fermion masses are hierarchical, according to
\ini
\frac{m_u}{m_c} \sim \frac{m_c}{m_t} \sim \lambda^4
\fin
\ini
\frac{m_d}{m_s} \sim \frac{m_s}{m_b} \sim \lambda^2
\fin
\ini
\frac{m_e}{m_{\mu}} \sim \frac{m_{\mu}}{m_{\tau}} \sim \lambda^2,
\fin
where $\lambda=0.22$ is the Cabibbo parameter. Hence, the charged lepton
and down quark mass hierarchies are similar.
We insert these hierarchies into the quark mass matrices of Ref.\cite{dv}
and then obtain six approximate forms for the Dirac neutrino mass matrix:
\ini
M_{\nu}^{I} \sim \left( \begin{array}{ccc}
0 & 0 & \lambda^4 \\ 0 & \lambda^4 & 0 \\ \lambda^4 & 0 & 1
\end{array} \right) m_t,
\fin
\ini
M_{\nu}^{II} \sim \left( \begin{array}{ccc}
0 & 0 & \lambda^4 \\ 0 & \lambda^4 & \lambda^4 \\ \lambda^4 & \lambda^4 & 1
\end{array} \right) m_t,
\fin
\ini
M_{\nu}^{III} \sim \left( \begin{array}{ccc}
0 & \lambda^8 & \lambda^4 \\ \lambda^8 & \lambda^4 & 0 \\ \lambda^4 & 0 & 1
\end{array} \right) m_t,
\fin
\ini
M_{\nu}^{IV} \sim \left( \begin{array}{ccc}
0 & \lambda^6 & \lambda^8 \\ \lambda^6 & 0 & \lambda^2 \\
\lambda^8 & \lambda^2 & 1
\end{array} \right) m_t,
\fin
\ini
M_{\nu}^{V} \sim \left( \begin{array}{ccc}
0 & \lambda^6 & 0 \\ \lambda^6 & \lambda^4 & \lambda^4 \\ 0 & \lambda^4 & 1
\end{array} \right) m_t,
\fin
\ini
M_{\nu}^{VI} \sim \left( \begin{array}{ccc}
0 & \lambda^3 & 0 \\ \lambda^3 & \lambda^2 & \lambda^2 \\ 0 & \lambda^2 & 1
\end{array} \right) m_b.
\fin
The charged lepton mass matrix is in the form (15) and gives negligible lepton
mixing. The matrix (10) contains three texture zeros. The other matrices contains
only two texture zeros.
We adopt such matrices to calculate the baryon asymmetry,
according to the formulas (2),(3). Since we are interested in the general trend,
without considering in detail the $CP$ violating phases,
we drop the imaginary part in Eqn.(3).
The required value for $Y_B$, from primordial nucleosynthesis, is in the range
$10^{-11}-10^{-10}$ (see for example the report \cite{osw}).

For the Dirac matrices $M_{\nu}^{I-III}$ we get
\ini
M_{R} \sim \left( \begin{array}{ccc}
\lambda^8 & \lambda^8 & \lambda^4 \\ \lambda^8 & \lambda^8 & \lambda^4 \\
\lambda^4 & \lambda^4 & 1
\end{array} \right) \frac{m_t^2}{m_1},
\fin
\ini
U_{R} \sim \left( \begin{array}{ccc}
1 & 1 & \lambda^4 \\ -1 & 1 & \lambda^4 \\
\lambda^4 & -\lambda^4 & 1
\end{array} \right),
\fin
and $y_D^{\dg} y_D=y_R$, with $y_R=M_R/v_R$.
The parameter $v_R \simeq M_3$ is the VEV of the singlet Higgs field which
generates the heavy neutrino mass (although it can be generated as bare
Majorana mass term). Note that in the 1-2 sector of $M_R$ all entries are of
the same order in $\lambda$. The eigenvalues of $M_R$ are
$M_3 \simeq m_t^2/m_1$, $M_2 \simeq \lambda^8 M_3$, and $M_1 \sim 0$
or $M_1 \sim M_2$. Here $M_1 \sim 0$ means that $M_1$ is much suppressed with
respect to $M_2$, and $M_1 \sim M_2$ means that they may differ by about
one order in $\lambda$.
In the first case $\epsilon_1$ is suppressed and,
as a consequence, $Y_B$ is much smaller than the required value.
In the other case
\ini
\epsilon_1 \simeq \frac{3}{16 \pi}
\left( \frac{\lambda^{16}}{\lambda^8} \cdot 1+\frac{\lambda^8}{\lambda^8}
\cdot \lambda^8 \right) \sim 10^{-7},
\fin
and $\tilde{m}_1 \simeq m_1$, so that $Y_B \sim 10^{-11}$. A sufficient level
of baryon asymmetry can be obtained, but $m_1$ has not to be less than $10^{-5}$ eV,
whereas we know that it is less than $10^{-2}$ eV.
A more precise expression for $M_{\nu}$, keeping the democratic form (5)
valid, leads to the case $M_1 \sim 0$ and hence to a suppression of $Y_B$.

For the Dirac matrix $M_{\nu}^{IV}$ we have
\ini
M_{R} \sim \left( \begin{array}{ccc}
\lambda^{12} & \lambda^8 & \lambda^6 \\ \lambda^8 & \lambda^4 & \lambda^2 \\
\lambda^6 & \lambda^2 & 1
\end{array} \right) \frac{m_t^2}{m_1},
\fin
with eigenvalues $M_3 \simeq m_t^2/m_1$, $M_2 \simeq \lambda^4 M_3$ and
$M_1 \simeq \lambda^{12} M_3$. We have also
\ini
U_{R} \sim \left( \begin{array}{ccc}
1 & \lambda^4 & \lambda^6 \\ -\lambda^4 & 1 & \lambda^2 \\
\lambda^6 & -\lambda^2 & 1
\end{array} \right),
\fin
and again $y_D^{\dg} y_D=y_R$. This relation is due to the democratic form (5).
The $CP$ violating asymmetry is 
\ini
\epsilon_1 \simeq \frac{3}{16 \pi}
\left( \frac{\lambda^{16}}{\lambda^{12}} \cdot \lambda^8+
\frac{\lambda^{12}}{\lambda^{12}}
\cdot \lambda^{12} \right) \sim 10^{-10},
\fin
and $\tilde{m}_1 \simeq m_1$, so that $Y_B \sim 10^{-14}$.
This is a too small value.

For the Dirac matrix $M_{\nu}^V$ we have
\ini
M_{R} \sim \left( \begin{array}{ccc}
\lambda^{12} & \lambda^{10} & \lambda^6 \\
\lambda^{10} & \lambda^8 & \lambda^4 \\
\lambda^6 & \lambda^4 & 1
\end{array} \right) \frac{m_t^2}{m_1},
\fin
with eigenvalues $M_3 \simeq m_t^2/m_1$, $M_2 \simeq \lambda^8 M_3$ and
$M_1 \simeq \lambda^{12} M_3$. We have also
\ini
U_{R} \sim \left( \begin{array}{ccc}
1 & \lambda^2 & \lambda^6 \\ -\lambda^2 & 1 & \lambda^4 \\
\lambda^6 & -\lambda^4 & 1
\end{array} \right),
\fin
and again $y_D^{\dg} y_D=y_R$. The $CP$ violating asymmetry is
\ini
\epsilon_1 \simeq \frac{3}{16 \pi}
\left( \frac{\lambda^{20}}{\lambda^{12}} \cdot \lambda^4+
\frac{\lambda^{12}}{\lambda^{12}}
\cdot \lambda^{12} \right) \sim 10^{-10},
\fin
and $\tilde{m}_1 \simeq m_1$, so that $Y_B \sim 10^{-14}$. Again this is a
too small value.

For the Dirac matrix $M_{\nu}^{VI}$ we find
\ini
\epsilon_1 \simeq \frac{3}{16 \pi}
\left( \frac{m_b}{m_t} \right)^2
\left( \frac{\lambda^{10}}{\lambda^{6}} \cdot \lambda^2+
\frac{\lambda^{6}}{\lambda^{6}}
\cdot \lambda^{6} \right) \sim 10^{-9},
\fin
$\tilde{m}_1 \simeq m_1$ and $Y_B \sim 10^{-13}$.
Even in this case the baryon asymmetry is too low.
Note that the matrix $M_{\nu}^{VI}$ can be obtained from
$M_{\nu}^V$ dividing powers of $\lambda$ by two and replacing $m_t$ with $m_b$.
The factor $(m_b/m_t)^2$ comes out from the different scale of Yukawa couplings.
It erases almost all the enhancement effect of the less pronounced hierarchy
of Yukawa couplings and of heavy neutrino masses. Instead, the dilution factor
is similar in any case, because of the relation $y_D^{\dg} y_D \simeq y_R$.
Dividing again powers of $\lambda$ by two, we find $\epsilon_1 \sim 10^{-7}$
and a sufficient amount of baryon asymmetry $Y_B \sim 10^{-11}$.

The previous results are valid in the nonsupersymmetric model. However, in the
supersymmetric model the amount of baryon asymmetry is only slightly
enhanced.
Model V with $M_R$ given by Eqn.(22) is similar to the one
considered in Ref.\cite{fal3}, based on the $U(2)$ horizontal symmetry, where a
different approximation for the dilution factor was used.
We confirm the results of the numerical analysis in Ref.\cite{ft}.

The relation $\tilde{m}_1 \simeq m_1$ has been found, under some
general circumstances, for the bilarge lepton mixing, in Ref.\cite{bs}.
With respect to that paper, we find $\epsilon_1$ and $Y_B$ smaller by
about three orders. This happens because assumptions A1 and A3, but not A2,
of Ref.\cite{bs} are fulfilled by matrices $M_{\nu}^{I-V}$. In particular,
some mixing angles in the unitary matrix that diagonalizes the Dirac neutrino
mass matrix are much smaller than the corresponding ratios of Dirac neutrino
masses, $s_{ij} \ll \sqrt{m_{\nu_i}/m_{\nu_j}}$.

\section{Conclusion}

In this paper we have studied the impact of the form of the Dirac neutrino mass
matrix $M_{\nu}$ on the generation of the baryon asymmetry through leptogenesis.
Assuming a full hierarchical spectrum of Majorana neutrinos, out of five
$M_{\nu}$ related to $M_u$ none of them generates a sufficient amount of
baryon asymmetry.
Even the matrix $M_{\nu}^{VI}$, related to $M_d$ (or $M_e$), is not able to
produce the required value of the asymmetry. If $M_1 \sim M_2$ the asymmetry
is enhanced, so that $M_{\nu}^{I-III}$ can produce enough
asymmetry. This case is not realized if a more precise expression for the
Dirac neutrino mass matrix is used. Therefore, we conclude that,
in the present context,
both the quark-lepton symmetry ($M_{\nu} \sim M_u$) and the up-down symmetry
($M_{\nu} \sim M_e$) for mass matrices are not reliable for the baryogenesis
via leptogenesis mechanism. We stress that, with a more moderate hierarchy in
$M_{\nu}$, we can get a sufficient amount of baryon asymmetry.

$~$

We thank M. Hirsch for communications.

\end{document}